\documentclass[twoside,slac_one]{revtex4}
\usepackage{graphicx}
\usepackage{subfigure}
\usepackage{fancyhdr}
\usepackage{amsmath} 
\usepackage{bm}
\usepackage{amsxtra}
\usepackage{amssymb}
\usepackage{amsthm}
\usepackage{latexsym}
\usepackage{lscape}

\begin{document}

\title{Measurement of the Top Pair Production Cross section at ATLAS}

%

\author{M. Saleem on behalf of the ATLAS collaboration}
\affiliation{Homer L. Dodge Department of Physics and Astronomy, University of Oklahoma, Norman, OK 73069, USA}

\begin{abstract}
We present the measurement of the production cross section for top quark pairs ($t\bar{t}$) in pp collisions at 
$\sqrt{s}$ = 7 TeV using the data recorded with the ATLAS detector at the Large 
Hadron Collider (LHC). Events are selected in single lepton (electron or muon) and dilepton ($ee$, $\mu\mu$, $e\mu$) topologies with 
multi-jets, and large missing transverse energy (MET). The combined result from these measurements is 
$\sigma_{t\bar{t}}$ = $176 \pm 5 \mathrm{(stat.)} ^{+13}_{-10}\mathrm{(syst.)} \pm 7 \mathrm{(lumi.)}  \mathrm{pb}$, which is 
in good agreement with the Standard Model prediction. 

\end{abstract}

\maketitle

\thispagestyle{fancy}


\section{Introduction}
The precision measurement of the top quark pair ($t\bar{t}$) production cross section is of central importance to the LHC physics program. 
At the LHC, the top quark production is dominated by the gluon-gluon fusion and with a smaller contribution from quark-quark annihilation. Uncertainties 
on the theoretical predictions for the top quark pair production are now less than 10\%, and comparison with experimental measurements performed 
in different channels allow a precision test of the predictions of perturbative Quantum Chromodynamics (QCD). Furthermore, top quark pair 
production is an important background in many searches for physics beyond the Standard Model (SM), and new physics may also give rise to additional $t\bar{t}$ 
production mechanisms or modifications of the top quark decay channels, which can affect the $t\bar{t}$ cross section. \\
All the analyses presented in this note were performed using the data collected by the ATLAS detector during the years 2010 and 2011. 
The single lepton channel results are based on 35 pb$^{-1}$ of data collected with the ATLAS detector~\cite{atlas} in 2010. Whereas the dilepton channel 
results are based on the 0.70 fb$^{-1}$ of data (partial data set) collected with the ATLAS detector in 2011. All these 
results were cross checked by independent analysis methods and only the methods that give the best performance are presented here. 

\section{Top quark pair production and decay}

Within the SM, the top quark pair production cross section in $pp$ collisions is calculated to be 165$^{+11}_{-16}$ pb~\cite{theory} at a center of 
mass energy at $\sqrt{s}$ = 7 TeV assuming a top mass of 172.5 GeV. 
Top quarks are predicted to almost always decay 100\% of the time to a W-boson and a b-quark ($t \rightarrow bW$). The decay topologies are 
classified according to the W-boson decays into a pair of quarks ($W \rightarrow q\bar{q'}$) or a lepton and a neutrino ($W \rightarrow l\nu$). 
When top quark pairs are produced, the branching ratios, for the single lepton mode with moderate background (where one of the W decays to a lepton and neutrino 
and the second W decays hadronically) 
and dilepton mode with low background (where both W's decay to a lepton and a neutrino), are 37.9\% and 6\% respectively. 
The corresponding final state signatures involve one or two leptons, jets, two of them are identified as the jets having b-quarks inside, and 
MET ($E_{T}^{miss}$), which is an indicator of undetected neutrino, from the W-boson decay. The single lepton channel has moderate 
background with the major contribution coming from W + jets and QCD multi-jets, where a jet is mis-identified as a lepton. The dilepton channel 
is very clean and has small background with the major contributions coming from Drell-Yan and QCD multi-jets.

\subsection{Object definition}
This section describes the general object definition used in top quark analyses. The reconstruction of $t\bar{t}$ events makes use of electrons, muons, 
jets, and MET. 
\begin{itemize}
\item {\bf Electrons}: Electron candidates are defined as good isolated
\footnote[1]{Isolation is defined as follows: $E_{T}$ deposited in the calorimeter cells within $\Delta$R $<$ 0.2 around the electrons is required to be less than 4 GeV. 
$E_{T}$ deposited in the calorimeter cells within $\Delta$R $<$ 0.3 around the muons is required to be less than 4 GeV, sum of p$_T$ of tracks 
within $\Delta$R $<$ 0.3 around muons is required to be less than 4 GeV, and it is required that there are no jets reconstructed 
with p$_T$ $>$ 20 GeV within $\Delta$R $<$ 0.4 around the muons.} electromagnetic clusters associated to well measured tracks. 
They are required to satisfy E$_{T} >$ 20 GeV and 
$\mid \eta \mid <$ 2.47 (the candidates between barrel to end cap 
regions were excluded [1.37, 1.52]), where $\eta$ is the pseudo-rapidity which is defined as $\eta = -ln[\tan(\frac{\theta}{2})]$. 
To suppress the background from photon conversions, the track is required to have an associated hit in the innermost pixel layer.
\item {\bf Muons}: Muon candidates are reconstructed from track segments in different layers of the muon spectrometer, combined with a 
charged track reconstructed in the inner tracking detector. They are required to satisfy $\mid \eta \mid <$ 2.5 and p$_{T} >$ 20 GeV. 
Muon candidates are also required to be isolated.
\item{\bf Jets}: Jets are reconstructed with the anti-$k_{t}$ algorithm with distance parameter 
\footnote[2]{radial distance in ($\eta$, $\phi$) space: $\Delta$R = $\sqrt{(\Delta\eta)^2 + (\Delta\phi)^2}$}$\Delta$R = 0.4 from energy clusters of adjacent calorimeter cells after 
noise subtraction~\cite{NS}, calibrated at the electromagnetic scale appropriate for the energy deposited by electrons or photons. These jets are 
calibrated to the hadronic energy scale, using a correction factor which depends on p$_{T}$ and $\mid \eta \mid$. If the closest object to an electron 
candidate (before the above electron isolation requirement) is a jet with a separation $\Delta$R $<$ 0.2 the jet was removed to avoid double-counting 
of electrons as jets. 
They are required to satisfy $\mid \eta \mid <$ 2.5 and p$_{T} >$ 25 GeV.
\item{\bf b-jets}: Jets arising from the hadronisation of b-quarks were identified using an algorithm (JetProb)~\cite{btagg} which relies upon the 
transverse impact parameter $d_{0}$ of the tracks in the jet: this is the distance of closest approach in the transverse $x$-$y$ plane of a 
track to the primary vertex (PV). On the basis of the impact parameter significance, $d_{0}/\sigma_{d_{0}}$, a probability that the 
track originates from the PV is computed. These individual track probabilities are then combined into a probability that the jet originates from 
the PV. Another method is based on the secondary vertex (SV) finding algorithm. The dilpeton channel uses a combination of the two 
algorithms based on the IP based taggers as well as the SV taggers. The combination of the two gives the higher $b$-tagging efficiency 
(80\%), which refers to the $b$-jets in $t\bar{t}$ events.
\item{\bf MET ($E_{T}^{miss}$)}: $E_{T}^{miss}$ is constructed from the vector sum of transverse calorimeter energy corrected for 
the reconstructed objects (jets, muons and electrons), such that the clusters included in reconstructed objects are calibrated 
according to the energy scale for that object. The $E_{T}^{miss}$ is required to be greater than 35 GeV.
\item{\bf Trigger}: It is required that an electron or a muon must have fired the single lepton trigger. The triggers used are fully efficient for 
leptons with p$_{T} >$ 20 GeV.
\end{itemize}
In addition to the selection listed above, contributions from non-collision backgrounds (e.g; beam-halo, cosmics) were also removed.  
For this the selected events were required to have at 
least one offline reconstructed PV with a minimum of five tracks. In the case of the dilepton channels, cosmic dimuon events were also vetoed. 

\subsection{Background Evaluation}
The main backgrounds to $t\bar{t}$ signal events in single lepton plus jets channel arise from W boson production in association with jets and from 
QCD multi-jet production. The later background can arise from various sources: 
\begin{itemize}
\item
jets can be mis-reconstructed as electrons if a relatively high 
fraction of their energy is in the electromagnetic calorimeter, 
\item
real electrons or muons can be produced in the decays of heavy flavor hadrons 
inside jets, or 
\item
photons inside jets can undergo conversions to electrons. 
\end{itemize}
Relatively smaller backgrounds also arise from Z+jets, diboson 
and single top production. The multi-jet background was measured with the data driven approach. The other, smaller, backgrounds have been 
estimated 
from MC simulations, normalized to the latest theoretical predictions. The W+jets background shape is determined from the MC while the normalization 
was extracted from the fit. The multi-jet background was measured (both shape and normalization) with data-driven 
approaches, commonly known as the matrix method for the muon channel and anti-electron method for the electron channel. \\
For the dilepton channels, the multi-jet background was extracted from data driven methods, similar to the single lepton channel. In this case 
the main background contribution is from the Drell-Yan production and it is suppressed by requiring same flavor events to have a cut on 
$E_{T}^{miss}$ (mentioned in section 3.2)  
and also imposing the Z boson mass window veto for $ee$ and $\mu\mu$ channels. For $e\mu$ events, a cut on the the scalar sum of jet 
and lepton transverse energies ($H_{T}$) is required (see section 3.2). The remaining Drell-Yan and multi-jet contributions were estimated via data driven methods. 
The other smaller backgrounds (like Z+jets, single top) in the dilepton channels were estimated from theory predictions.

\section{Cross section ($\sigma_{t\bar{t}}$) measurements}
\subsection{Single lepton channel}
This analysis is performed using the 35 pb$^{-1}$ of data collected with the ATLAS detector during the year 2010. 
In the single lepton channel two complementary measurements were performed. Both measurements exploit the difference in kinematic 
distributions of signal and background events. A projective Likelihood (LH) is used to separate signal from the background. LH discriminant 
is constructed using the multiple variables described below. Monte Carlo (MC) signal and backgrounds models these variables very well for building the LH 
discriminant. Finally a fit to the LH discriminant in data is performed by the sum of the two templates (signal and background) taken from MC. 
The fit extracts the number of $t\bar{t}$ and W + jets events, which are then used to extract the cross section.\\
For the first method, no explicit identification of
secondary vertices inside jets (no $b$-tagging requirement) is performed. The $t\bar{t}$ production cross section was extracted by exploiting 
the different kinematic properties of the $t\bar{t}$ 
events with respect to the dominant W+jets background. Three variables are selected for their discriminant power, optimized for the small 
correlation between them and also by considering the effect of the jet energy scale (JES) as well as statistical uncertainties. 
The variables are:
\begin{itemize}
\item the event aplanarity (to smooth the aplanarity distribution $e^{-8\times\mathcal{A}}$ is used), exploits the fact that $t\bar{t}$ events are more isotropic than W+jets events,
\item the lepton pseudo-rapidity ($\eta_{lepton}$), exploits the fact that $t\bar{t}$ events produced have more central leptons than W+jets events,
\item the charge of the lepton ($q_{lepton}$), exploits the fact that $t\bar{t}$ events produce charge symmetric leptons while the W+jets events 
produce asymmetric leptons.
\end{itemize}
We built a likelihood (LH) discriminant from the above input variables, following the projective LH approach using the TMVA package~\cite{tmva}. 
Two jet multiplicity bins are considered: exactly 3jets (lepton + 3jets) and $\geq$ 4jets
(lepton + $\geq$ 4jets). Individual LH functions are defined for each of the 4 channels and multiplied together in a combined fit to extract the 
total number of signal events (N$_{t\bar{t}}$). The performance of the LH fit is estimated by performing pseudo-experiments. The distributions of 
the three input variables (mentioned above) and of the LH discriminant in data and simulated events are in good agreement.
The major systematic uncertainties on this measurement are the amount of initial and final state radiation, as well as limited understanding of the 
JES and reconstruction efficiency.
The systematic 
uncertainties associated with the simulation, object definitions and the QCD multi-jet estimate, as well as the statistical uncertainty and 
the uncertainty on the luminosity are summarized in the Table~\ref{tab:mult_syst}. 
A cross section extracted after the fit to the combined 
LH discriminant is given as follows: 
\[
{\sigma}_{t{\overline t}} = 171 \pm 17 \mathrm{(stat.)} ^{+20}_{-17}\mathrm{(syst.)} \pm 6 \mathrm{(lumi.)} \mathrm{\ pb}.
\]
The measurement agrees with approximate NNLO perturbative QCD
calculations. Cross-check measurements are performed with
one-dimensional likelihood fits and ``cut-and-count'' methods which are
found to be consistent with the main result~\cite{SL-nobtag}.

\begin{figure}[h]
\centering
{\includegraphics[width=135mm]{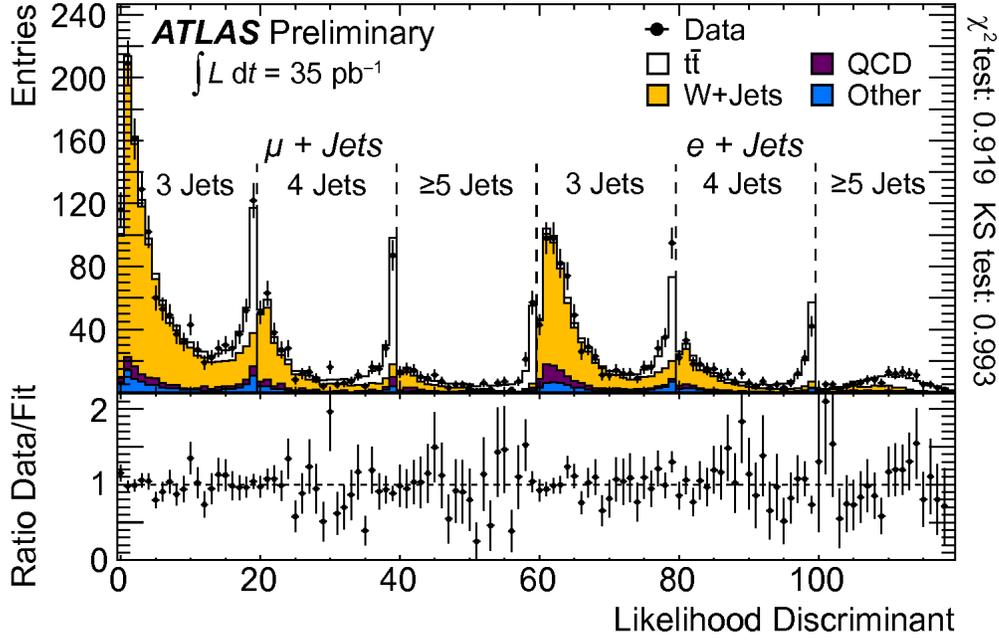}} 
\caption{Distribution of the likelihood discriminant (D) for the baseline analysis. Data are superimposed on expectations, scaled to the 
results of the fit. The left bins correspond to the muon channel and the right bins to the electron channel. The bottom plot shows the ratio of data to fit result.}
\label{fig:btagFit}
\end{figure}

\begin{figure}[h]
\centering
\subfigure[single lepton+jets channel without $b$-tagging]{\includegraphics[width=80mm]{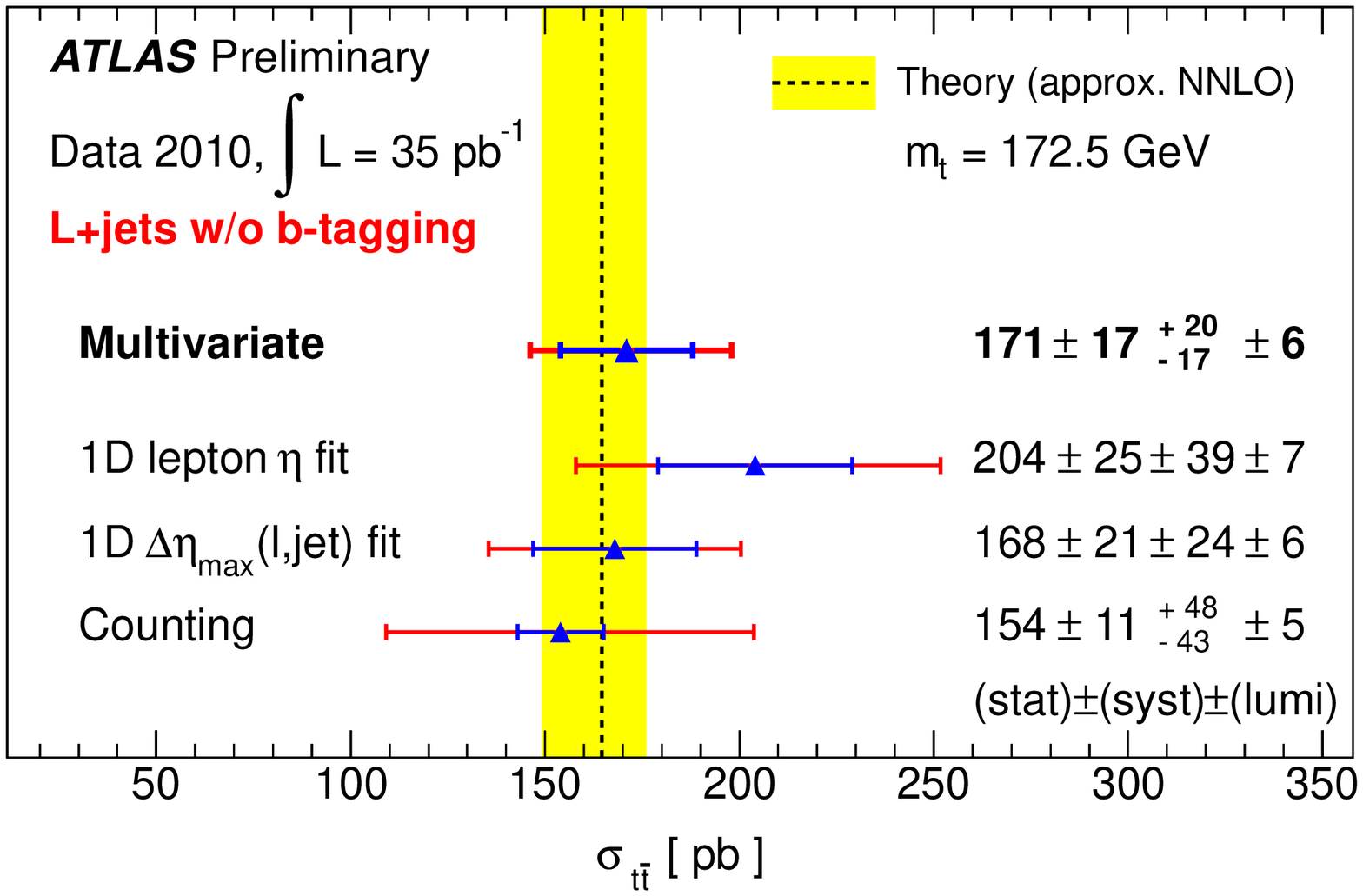}} 
\subfigure[single lepton+jets channel with $b$-tagging]{\includegraphics[width=80mm]{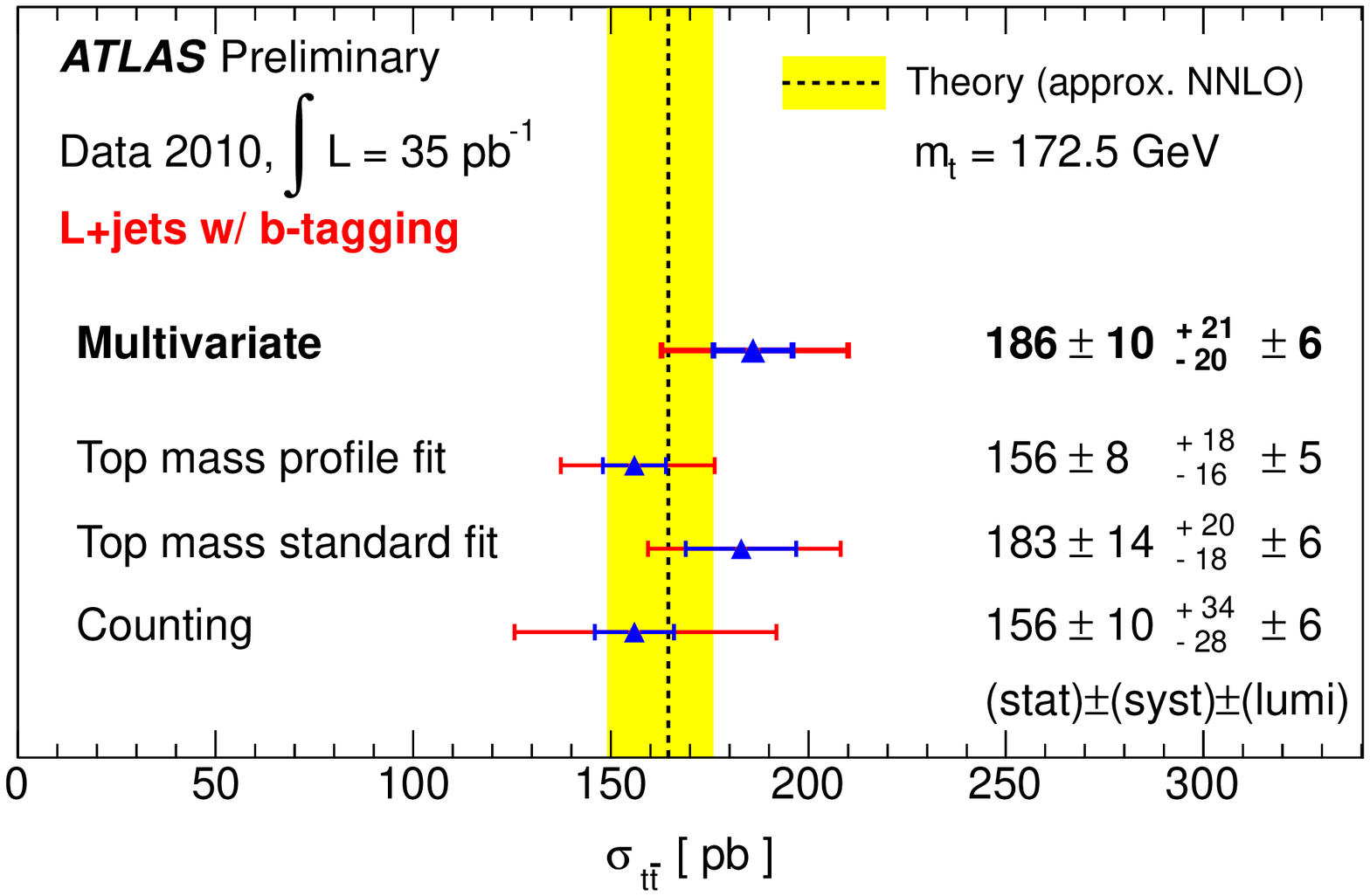}}
\caption{Summary of the $t\bar{t}$ cross section measurements in the single lepton+jets channel without (on left) and with (on right) $b$-tagging using 35 pb$^{-1}$ 
of data. The yellow band shows the approximate NNLO perturbative QCD prediction~\cite{theory}.}
\label{fig:results}
\end{figure}

\begin{table}[h]
\begin{center}
\caption{Summary of individual systematic
uncertainty contributions to the multivariate fit lepton+jets analysis without $b$-tagging.
}
\begin{tabular}{| l | c |}
\hline
\textbf{Source}   & \textbf{Relative cross-section uncertainty [\%]}   \\ \hline\hline
{\em Object selection}  &  \\
Lepton reconstruction, identification, trigger  &  -1.9 / +2.6         \\
Jet energy scale and reconstruction         &  -6.1 / +5.7          \\
\hline
{\em Background rates and shape}  & \\
 QCD normalization               &  $\pm3.9$       \\
 QCD shape               &  $\pm$3.4       \\
 W+jets shape               &  $\pm$1.2       \\
 Other backgrounds normalization         &  $\pm0.5$       \\
\hline
{\em Simulation}      & \\
Initial/final state radiation  & -2.1 / +6.1    \\
Parton distribution functions         &   -3.0 / +2.8       \\
Parton shower and hadronisation &   $\pm3.3$           \\
Next-to-leading-order generator                &   $\pm2.1$         \\
MC statistics                                   & $\pm 1.8$ \\
 Pile-up                      &   $\pm$1.2             \\
\hline
Total systematic uncertainty  & -10.2 / +11.6        \\
\hline
\end{tabular}
\label{tab:mult_syst}
\end{center}
\end{table}
A second method, in the single lepton channel, exploits $b$-tagging information in the context of a multivariate LH discriminant. Four variables 
are used to construct 
the multivariate LH distribution, among which the average of the weights of the two most significant $b$-tags (for the jets with lowest 
probability to originate from the PV). The other variables are lepton pseudo-rapidity, the aplanarity and $H_{T,3p}$ (transverse energy of 
all jets except the two leading ones, normalized to the sum of absolute values of all longitudinal momenta in the event). More channels are 
used here with an additional jet multiplicity bin (3, 4, and $\ge$ 5 jets). 
A profile LH fit with 17 nuisance parameters combining the six 
channels is performed to extract the $\sigma_{t\bar{t}}$ and constrain the effect of systematic uncertainties using data. Figure~\ref{fig:btagFit} 
shows the LH discriminant distribution for the selected data superimposed on the prediction. The fitter treats the templates of the six 
analysis channels with 20 bins each as one large 120-bin histogram, the left bins corresponding to the muon channel and the right bins to the 
electron channel. The expected contributions have been scaled according to the results of the fit. 
A cross section is extracted:
\[
{\sigma}_{t{\overline t}} = 186 \pm 10 \mathrm{(stat.)} ^{+21}_{-20}\mathrm{(syst.)} \pm 6 \mathrm{(lumi.)} \mathrm{\ pb}.
\]
Here the main systematic uncertainties originate from the $b$-tagging algorithm calibration from data and heavy flavor fraction in W+jets events.
This measurement also agrees with approximate NNLO perturbative QCD
calculations. The cross-check measurements are also performed with two
one-dimensional kinematic fits to the reconstructed top mass and a ``cut-and-count'' method which are
found to be in good agreement with the main result~\cite{SL-btag}. \\
The comparison of the results for the two measurements is shown in Figure~\ref{fig:results}.
\subsection{Dilepton channel}
For the dilepton case, the analysis are performed using 0.70 fb$^{-1}$ of data collected with the ATLAS detector.
In the dilepton channels the cross section is measured using a cut-and-count method both with and without requiring the $b$-tagging. Events 
are selected by requiring exactly two opposite signed leptons in the three channels, $ee$, $\mu\mu$, and $e\mu$. In this case, the major 
background contribution is from the Drell-Yan (DY) production. This is suppressed by (in the case of $ee$, $\mu\mu$) requiring same 
flavor events to have $E_{T}^{miss} >$ (60)40 GeV for the analyses (with) without $b$-tagging. 
For same sign, lepton modes, a Z-boson mass veto is also required by excluding events with $\mid$ $m_{ll} - m_{Z} \mid$ $<$ 10 GeV.   
\begin{figure}[h]
\centering
\subfigure[Jet multiplicity distribution for $ee + \mu\mu + e\mu$ events without $b$-tag]{\includegraphics[width=80mm]{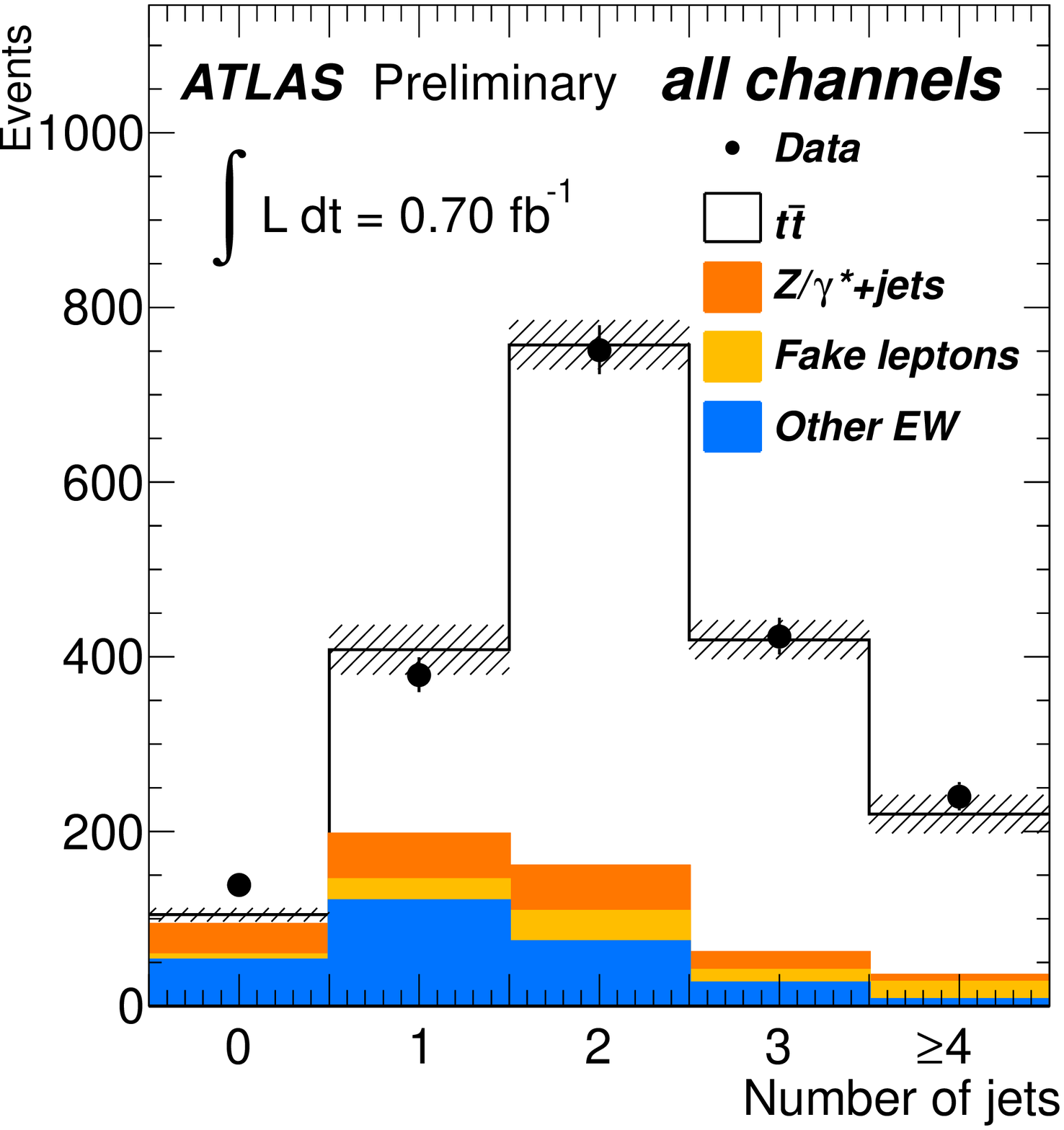}}
\subfigure[Multiplicity distribution of $b$-tagged jets in $ee + \mu\mu + e\mu$ events]{\includegraphics[width=80mm]{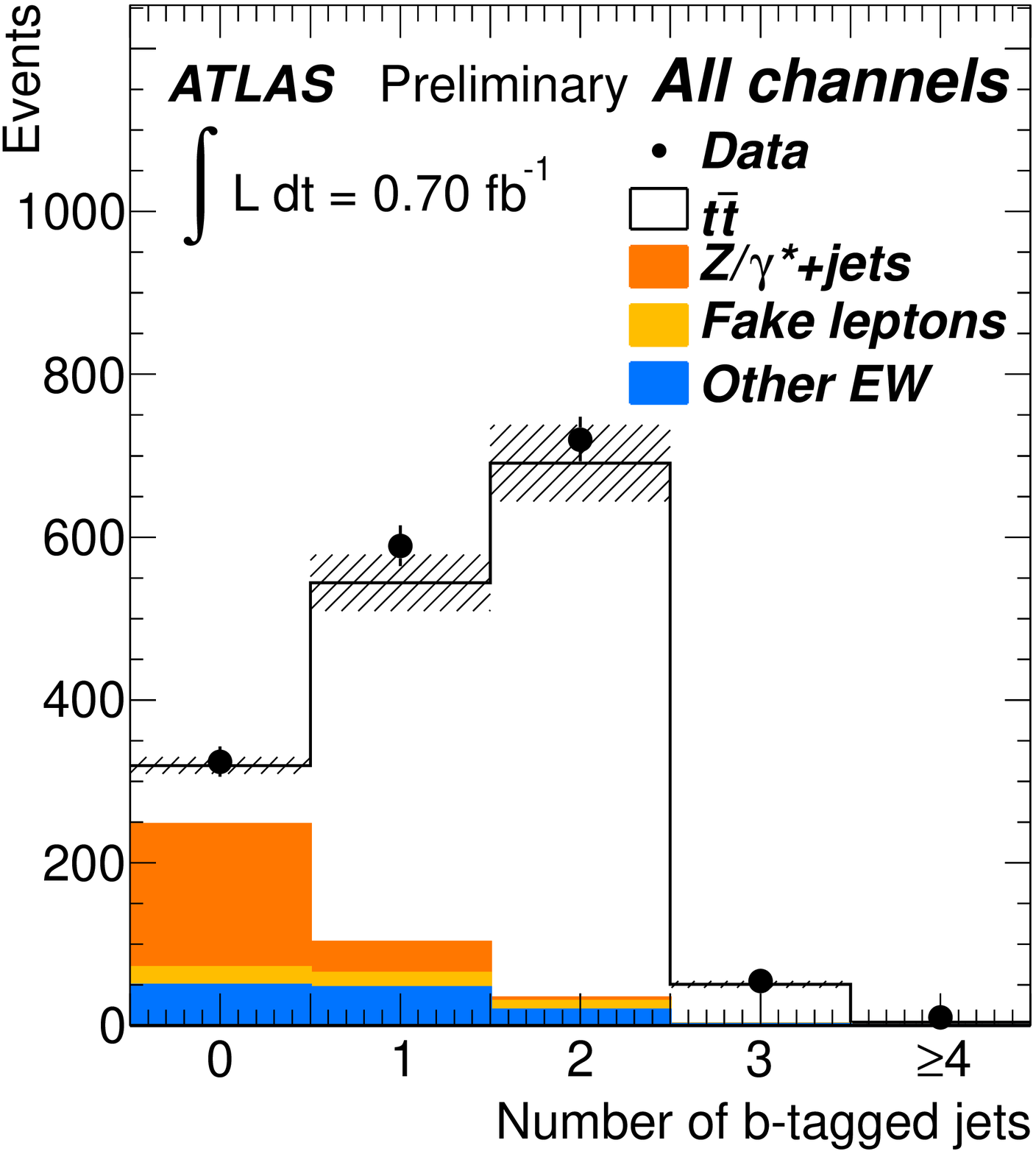}}
\caption{(a) Jet multiplicity distribution for $ee + \mu\mu + e\mu$ events without $b$-tag. (b) Multiplicity distribution of 
$b$-tagged jets in $ee + \mu\mu + e\mu$ events. Contributions from diboson and single top-quark events are summarized as "Other EW". 
Note that the events in (b) are not a simple subset of those in (a) because the event selections for the $b$-tag and non-$b$-tag 
analyses differ.} 
\label{fig:JetMul-fig}
\end{figure}
\begin{figure}[h]
\centering
\subfigure[$H_T$ distribution in $e\mu$ events without $b$-tag]{\includegraphics[width=80mm]{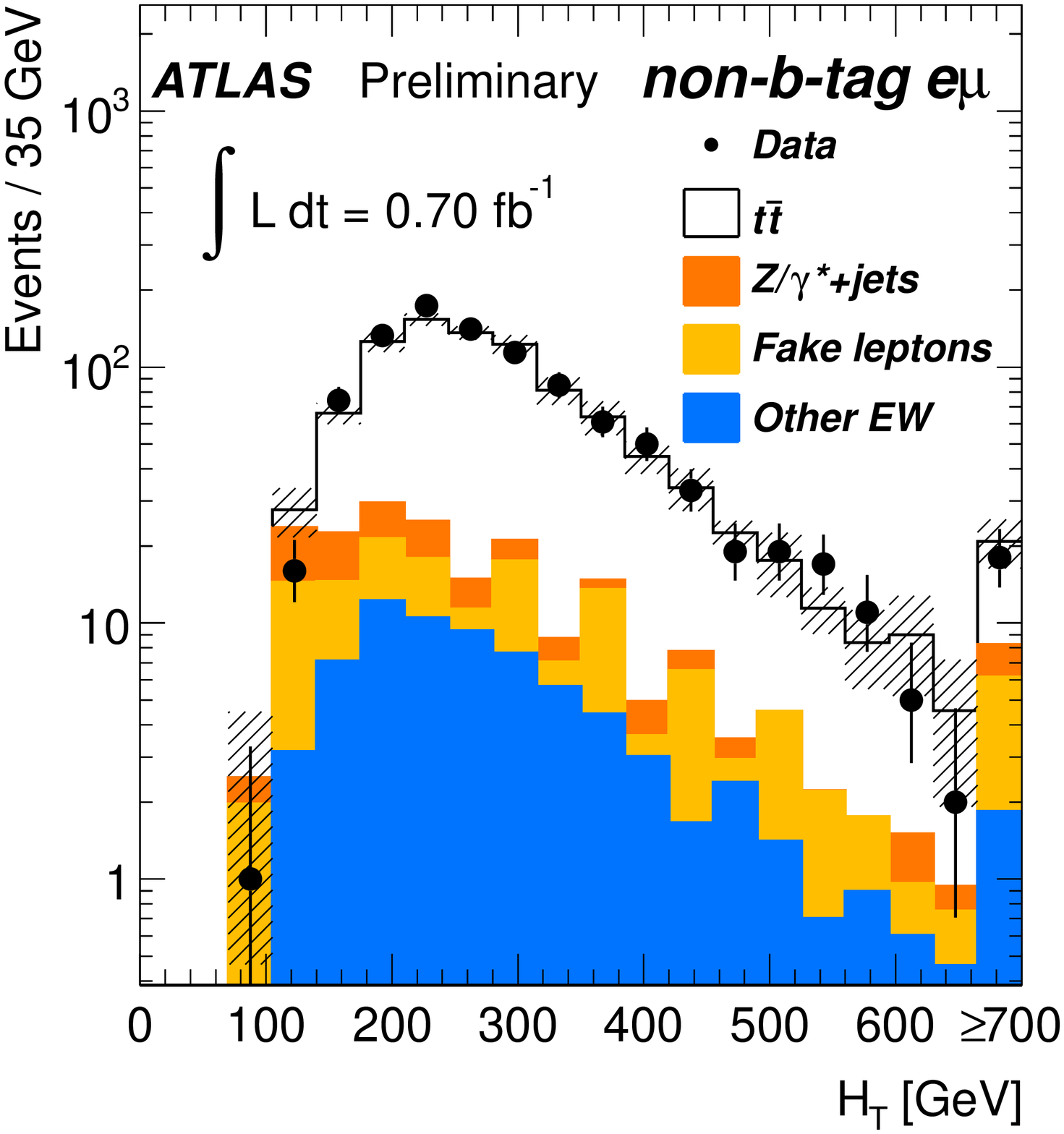}}
\subfigure[$H_T$ distribution of $b$-tagged $e\mu$ events]{\includegraphics[width=80mm]{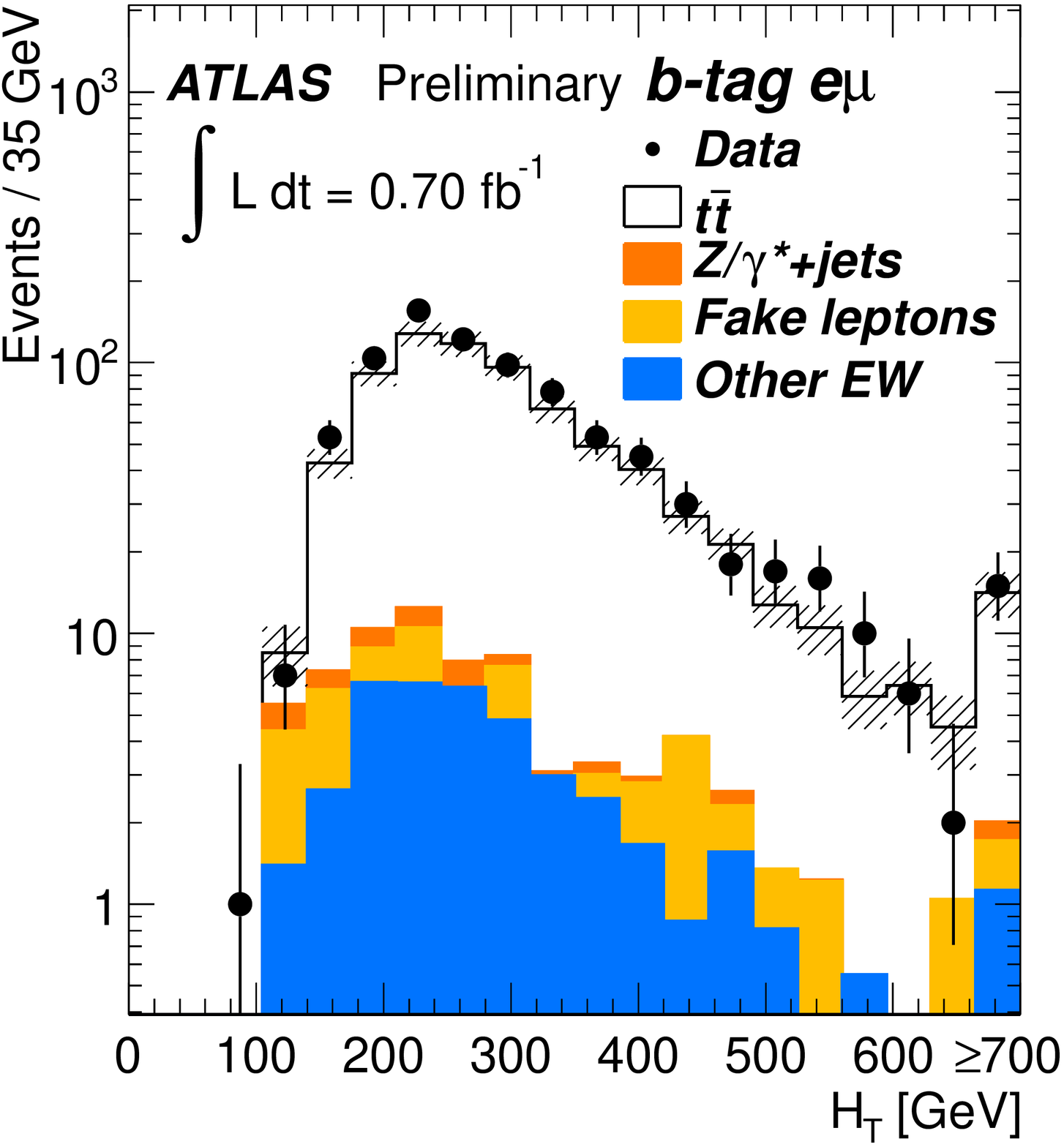}}
\caption{$H_{T}$ distribution in the signal region for the non-$b$-tag $e\mu$ channel (left) and with at least one $b$-tagged jet (right). 
Contributions from diboson and single top-quark events are summarized as "Other EW".} \label{fig:Ht-fig}
\end{figure}
\begin{figure}[h]
\centering
\subfigure[Single lepton and dilepton results comparison, with yellow line NNLO prediction]{\includegraphics[width=80mm]{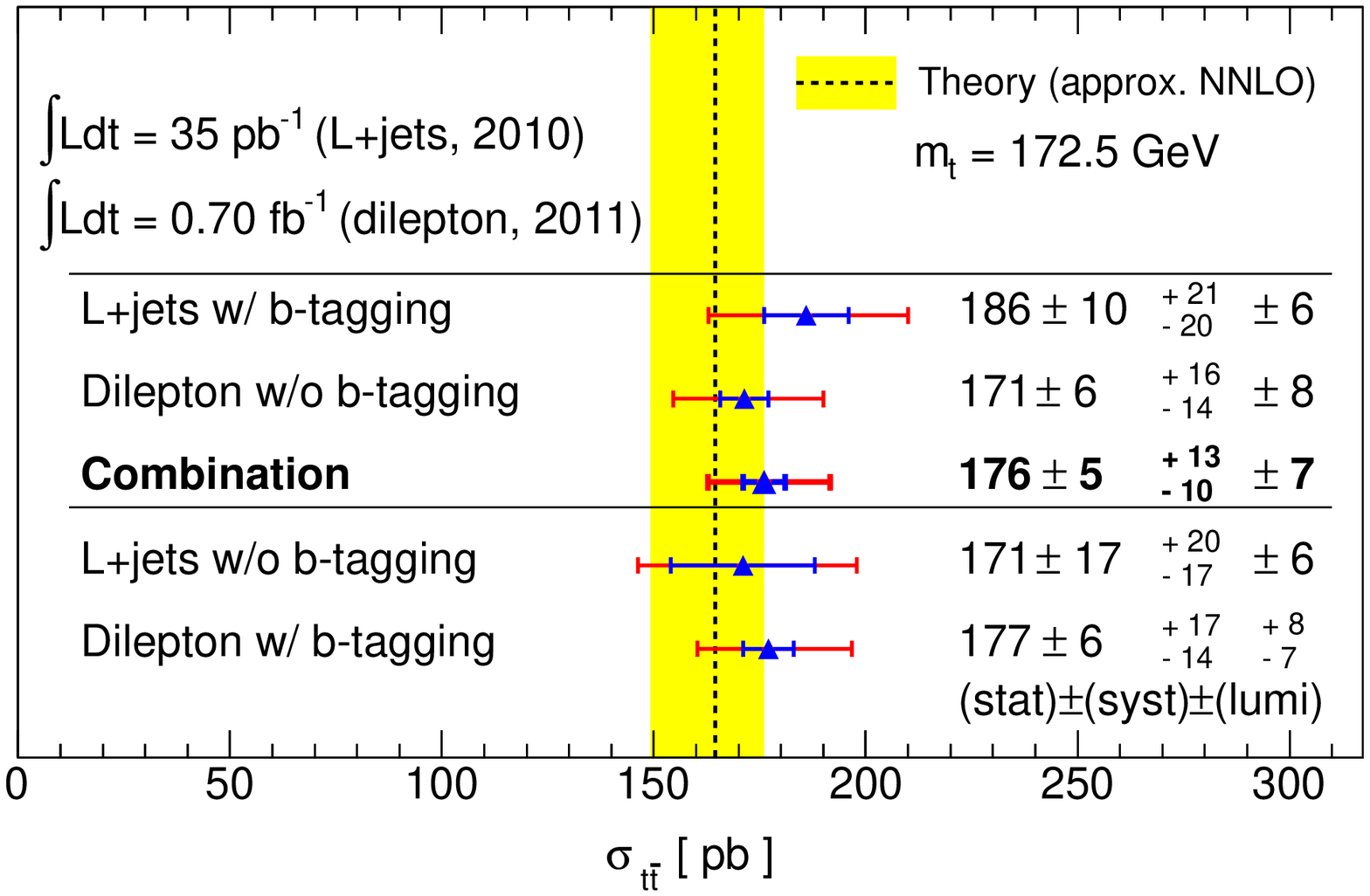}}
\subfigure[$\sigma_{t\bar{t}}$ on $\sqrt{s}$ = 7 TeV theory prediction and ATLAS measurement]{\includegraphics[width=80mm]{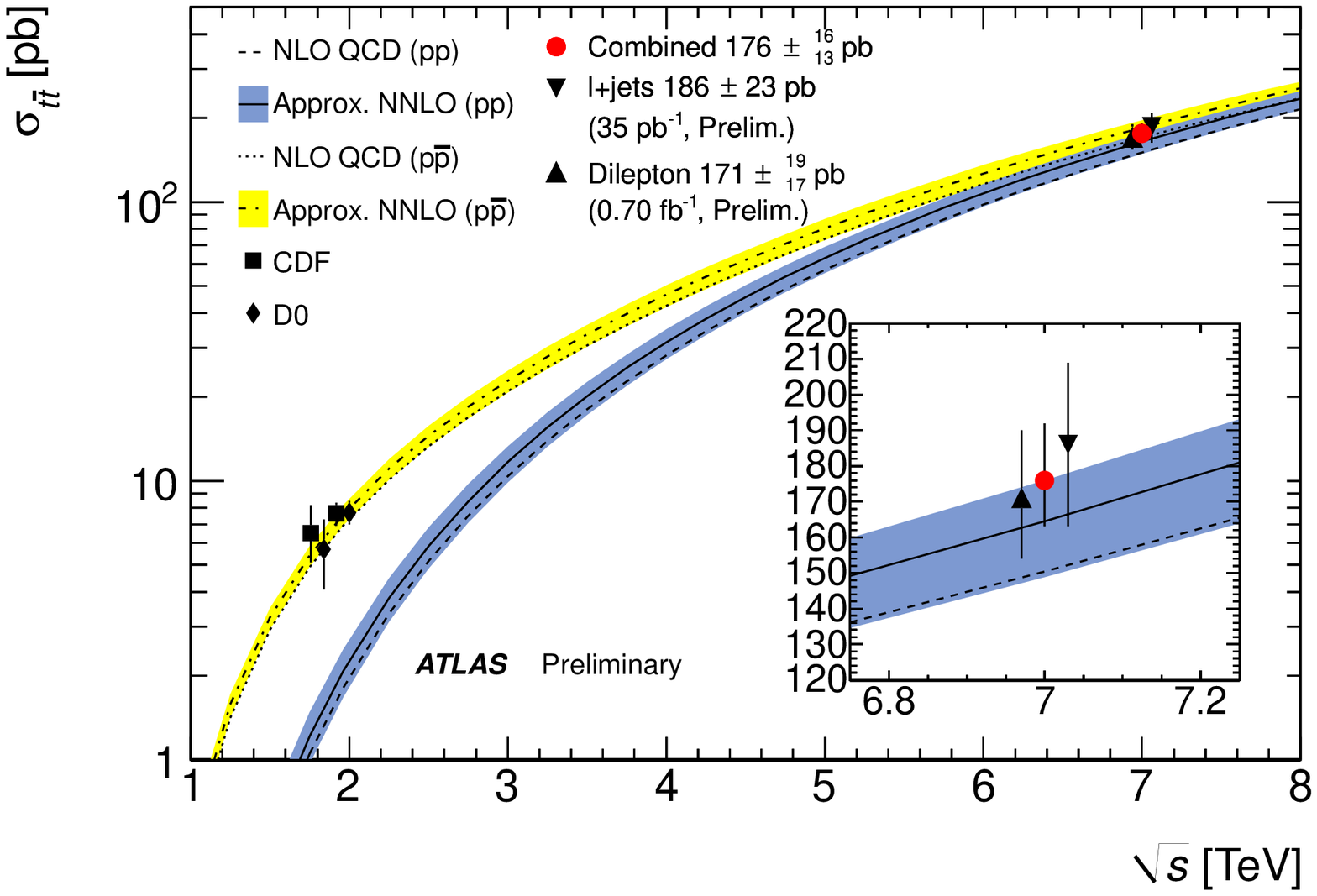}}
\caption{Plot (on left) of the measured value of $\sigma_{t\bar{t}}$ in the single-lepton with $b$-tagging channel, the dilepton without $b$-tagging 
channel, and the combination of these two channels, including error bars for both statistical uncertainties only (blue) and with full 
systematics (red). Results from auxiliary single-lepton and dilepton measurements are also shown. 
The approximate NNLO prediction is shown as a vertical dotted line with its error in yellow. 
The plot (on right) shows measurements of $\sigma_{t\bar{t}}$ from ATLAS in proton-proton collisions, 
and CDF and D0 in proton-antiproton collisions, compared to theoretical predictions assuming a top mass of 172.5 GeV as a function 
of $\sqrt{s}$. The present result is indicated by the red circle.}\label{fig:comb-fig}
\end{figure}
For $e\mu$ events, the scalar sum of jet and lepton transverse
energies ($H_T$) is required to be greater than 130(140) GeV for the analyses (with) without $b$-tagging. The remaining DY and fake lepton 
contributions are estimated using data driven techniques. The other SM background rates are estimated using the predictions from simulation. 
In Figure~\ref{fig:JetMul-fig} the numbers of selected jets and the 
expectation for 0.70 fb$^{-1}$ are shown for the non-$b$-tag analysis, and for the $b$-tag analysis, with the three channels combined. 
In the non-$b$-tag case, all requirements except the jet multiplicity selection are applied, and in the $b$-tag case all requirements
except the $b$-tag requirement are applied. The distributions of $H_T$ are shown in Figure~\ref{fig:Ht-fig} for the $e\mu$ channel 
in both $b$-tagged and non-$b$-tagged cases. All requirements except $H_T$ are applied.
The cross section from the three channels is combined using a profile LH with a simultaneous fit to them. The extracted combined cross section 
for the analysis without $b$-tagging is:
\[
{\sigma}_{t{\overline t}} = 171 \pm 6 \mathrm{(stat.)}  ^{+16}_{-14}\mathrm{(syst.)} \pm 8 \mathrm{(lumi.)} \mathrm{\ pb}.
\]
A second measurement requiring at least one jet consistent with a $b$ quark yields: 
\[
{\sigma}_{t{\overline t}} = 177 \pm 6 \mathrm{(stat.)}  ^{+17}_{-14}\mathrm{(syst.)} ^{+8}_{-7}\mathrm{(lumi.)} \mathrm{\ pb}.
\]
These measurements are in good agreement with each other and with Standard Model predictions~\cite{dilepton}.
\section{Combination}
The two most precise cross section measurements in the dilepton and single lepton are combined as shown in Figure~\ref{fig:comb-fig}. 
The combination uses the dilepton cut-and-count without $b$-tagging and single lepton with $b$-tagging. The combined result is based 
on the estimation of a profile LH ratio and taking into account the correlations in the systematic uncertainties~\cite{comb}. The total 
uncertainty (including systematics and statistics) on the combined cross section comes out to be 10\%. 
\[
{\sigma}_{t{\overline t}} = 176 \pm 5 \mathrm{(stat.)}  ^{+13}_{-10}\mathrm{(syst.)} \pm 7 \mathrm{(lumi.)} \mathrm{\ pb}.
\] 
This is in good agreement with 
the SM prediction as shown in figure~\ref{fig:comb-fig}.
\section{Results and Conclusion}
The excellent performance of the LHC has provided us with enough data that many of these measurements are now limited by systematic uncertainties.    
The precision on the ATLAS top quark pair production measurements are comparable with the uncertainty of the NNLO prediction. 
In conclusion, the ATLAS top quark production measurements are entering an era of precision.
%
%
\clearpage
\bigskip 
\begin{acknowledgments}
I wish to thank all the LHC teams and the ATLAS Collaboration for the wonderful performances of the  accelerator complex and the 
excellent data quality, the many authors and contributors of top physics analyses for reading and commenting this manuscript. I also 
gratefully acknowledge the support of Department of Energy (USA) and of my institue for funding this research.

\end{acknowledgments}
\bigskip 

\begin{thebibliography}{99} 

\bibitem{atlas} The ATLAS Collaboration, Eur. Phys. J. {\bf C 71} (2011)1577. 
\bibitem{theory} M.~Aliev {\it et al.}, Hadronic Top and Heavy  quarks cross section calculator, arXiv:1007.1327 [hep-ph].
\bibitem{NS} The ATLAS Collaboration, Measurement of inclusive jet and dijet cross-sections in proton-proton collisions at 7 TeV centre-
of-mass energy with the ATLAS detector, Eur. Phys. J. {\bf C 71} (2011).
\bibitem{btagg} The ATLAS Collaboration, Calibrating the b-tag Efficiency and Mistag Rate in 35 pb$^{-1}$ of Data with the Atlas Detector, ATLAS-CONF-2011-089 http://cdsweb.cern.ch/record/1356198.
\bibitem{tmva} A. Hoecker, P. Speckayer, J. Stelzer, J. Therhaag, E. von Toerne, H. Voss, Toolkit for Multivariate Data Analysis with ROOT, arXiv:physics/0703039 (2007).
\bibitem{SL-nobtag} The ATLAS Collaboration, Top Quark Pair Production Cross section Measuremnets in ATLAS in the single Lepton+Jets Channel without $b$-tagging, ATLAS-CONF-2011-023 http://cdsweb.cern.ch/record/1336753.
\bibitem{SL-btag} The ATLAS Collaboration, Top Quark Pair Production Cross section Measuremnets in ATLAS in the single Lepton+Jets Channel with $b$-tagging, ATLAS-CONF-2011-035 http://cdsweb.cern.ch/record/1337785.
\bibitem{dilepton} The ATLAS Collaboration, Measurement of the top-quark pair production cross-section in pp collisions at $\sqrt{s}$ = 7 TeV in dilepton final states with ATLAS, ATLAS-CONF-2011-100 http://cdsweb.cern.ch/record/1369215.
\bibitem{comb} The ATLAS Collaboration, Measurement of the top quark pair production cross-section based on a statistical combination of measurements of dilepton and single-lepton final states 
at $\sqrt{s}$ = 7 TeV with the ATLAS detector, ATLAS-CONF-2011-108 http://cdsweb.cern.ch/record/1373410.


%

\end{thebibliography}

\end{document}